\documentclass[conference]{IEEEtran}
\IEEEoverridecommandlockouts
\pdfoutput=1
\usepackage{cite}
\usepackage{multirow}
\usepackage{booktabs}
\usepackage{amsmath,amssymb,amsfonts}
\usepackage{algorithmic}
\usepackage{graphicx}
\usepackage{textcomp}
\usepackage{xcolor}
\usepackage{float}

\usepackage{fancyhdr}
\def\BibTeX{{\rm B\kern-.05em{\sc i\kern-.025em b}\kern-.08em
    T\kern-.1667em\lower.7ex\hbox{E}\kern-.125em}}

\begin{document}

\title{Large-Scale Knowledge Synthesis and Complex Information Retrieval from Biomedical Documents\\
\thanks{\textsuperscript{*} Equal contribution, \textsuperscript{+} Corresponding author}
\thanks{Shorter version of this work is published at IEEE BigData 2022 Conference, held at Osaka, Japan \cite{10020725} DOI:10.1109/BigData55660.2022.10020725}
}


\author{\IEEEauthorblockN { Shreya Saxena\textsuperscript{*} 
, Raj Sangani\textsuperscript{*}, Siva Prasad\textsuperscript{*}, Shubham Kumar\textsuperscript{*}} {Mihir Athale\textsuperscript{*}, Rohan Awhad\textsuperscript{*}, Vishal Vaddina\textsuperscript{+}}\\
\IEEEauthorblockA{\textit{Applied Research, Quantiphi}\\
\{shreya.saxena, siva.prasad, shubham.kumar01, mihir.athale, rohan.awhad, vishal.vaddina\}@quantiphi.com}
rsangani@ucdavis.edu
}

\maketitle

\begin{abstract}
Recent advances in the healthcare industry have led to an abundance of unstructured data, making it challenging to perform tasks such as efficient and accurate information retrieval at scale. Our work offers an all-in-one scalable solution for extracting and exploring complex information from large-scale research documents, which would otherwise be tedious. First, we briefly explain our knowledge synthesis process to extract helpful information from unstructured text data of research documents. Then, on top of the knowledge extracted from the documents, we perform complex information retrieval using three major components- Paragraph Retrieval, Triplet Retrieval from Knowledge Graphs, and Complex Question Answering (QA). These components combine lexical and semantic-based methods to retrieve paragraphs and triplets and perform faceted refinement for filtering these search results. The complexity of biomedical queries and documents necessitates using a QA system capable of handling queries more complex than factoid queries, which we evaluate qualitatively on the COVID-19 Open Research Dataset (CORD-19) to demonstrate the effectiveness and value-add. 
\\
\end{abstract}
\begin{IEEEkeywords}
Information-Retrieval, Knowledge-Synthesis, Semantic-Retrieval, Question-Answering, CORD-19
\end{IEEEkeywords}

\section{Introduction}
The healthcare sector stands tall with an enormous amount of unstructured text data in documents, articles, biomedical journals, and JSON files, as well as structured data like tables, and electronic health records, often leading to a severe information overload challenge. To researchers and health professionals, extracting relevant information from a huge corpus of biomedical data is a complex and tedious task that delays the research outcome and involves considerable capital. Therefore, there is an increased urgency for an information retrieval system in the biomedical domain to retrieve such complex information.
\\
Information Retrieval (IR) is a core task in many real-world applications, such as digital libraries, expert finding, web search and others. Information retrieval aims at retrieving information relevant to a query from large data collections, which has been an active research area in the healthcare domain \cite{sakji2009information}\cite{al2020information} \cite{weiming2008knowledge}. Traditional information retrieval systems rely on lexical retrievers such as Boolean Retrieval, BM25 \cite{jones2000probabilistic}, and statistical language models, which aim to find an exact match between the query and documents but fail to handle the problem of vocabulary and semantic mismatch. Earlier studies in neural IR handle the problem of vocabulary mismatch by taking a different approach, such as maximum inner product search (MIPS) between GLoVe \cite{pennington2014glove} or Word2vec \cite{mikolov2013efficient} embeddings of query and document terms. The problem of semantic mismatch was solved by leveraging contextual embeddings with the introduction of language models \cite{devlin2018bert}. Lexical systems might fail to capture the semantics of the concepts, especially in biomedical data with complex terms that sometimes are quite ambiguous. Semantic systems can handle this ambiguity well, but these systems often have difficulty dealing with longer contexts. Hence, we need a hybrid framework that can accommodate both of these mechanisms. This paper conceptualizes a framework to help users access meaningful information extracted from massive corpora in the biomedical domain. They can explore the information in the form of knowledge graphs (Section III.A), search for specific information, and get answers to complex questions i.e, questions that require multiple contexts to provide an answer (e.g. \textit{“What virus was isolated from a patient who died from acute respiratory failure?”}).
\\
We propose an all-in-one information retrieval framework using lexical and semantic approaches, shown in Fig.1, that combines multiple functionalities like passage retrieval; triplet retrieval from knowledge graphs and complex QA. We also include faceted navigation for filtering the triplet search results, making it easier for the user to explore relevant information through a large amount of data. Our question answering system can answer complex queries by integrating multi-hop dense retriever \cite{xiong2020answering}, which uses a dense iterative retrieval method.
\\
The following describes how the paper is structured: Background information is provided in Section-II. The methodology is then discussed in Section-III with distinct subsections for its various components, experiments are discussed in Section-IV, and Section-V concludes the paper.

\section{Background}
To researchers and health professionals, extracting relevant information from a huge corpus of medical research documents and texts is a complex, time-consuming, and tedious task that delays the research outcome and involves considerable capital, yet is a necessity. Therefore, there has been an increased urgency for an information retrieval system in the medical domain to retrieve such complex information \cite{article}.
Recent years have witnessed  an increase in information retrieval systems in the healthcare domain, such as a medical information retrieval system for e-healthcare records \cite{sengan2020medical}, retrieval of semantically similar questions in healthcare forums \cite{wang2015retrieval}, and a system that uses information retrieval with an added component for classifying breast cancer \cite{kumari2022intelligent}. 
\\
Due to the pandemic, information extraction around COVID-19 data has emerged as an active research area \cite{shorten2021deep}, predominantly using knowledge graphs \cite{wise2020covid}\cite{esteva2021covid}. Complex question answering, especially in the medical domain, has also become prominent \cite{jin2022biomedical}. These systems try to solve problems like knowledge graph (KG) generation on structured data, factoid question answering and searching entities in the KG \cite{lan2021complex}. These systems fail to address these use cases comprehensively. For example, the KG created might be ontology specific and cannot capture facts from open text or might only represent metadata information in the form of a graph; these systems also fail to provide an efficient integrated search \cite{su2020caire}\cite{ambavi2020covidexplorer} and QA functionality such as ours. 
\\
Recent transformer-based retrievers mostly rely on the maximum inner product search between the dense representation of the query and the documents, generated using transformer models. These retriever-based systems are often supported by a re-ranker, based on variants of transformer models like SBERT and BERT-based cross-encoders \cite{reimers2019sentence}. 
\\
Previous work on Open Domain Question Answering \cite{zhu2021retrieving} is mainly based on retriever and reader architecture, Iterative Retriever, Reader, Reranker (IRRR) \cite{qi2020answering}, which captures an initial set of keywords from the query, expands it based on the passages it retrieves from the database and re-ranks them. These re-ranked passages are then passed to the reader to generate answers, and the whole process is iteratively repeated until the answer is found with high confidence. All of this makes the IRRR based systems highly complex due to the large number of components involved, coupled with longer inference time and higher memory consumption. Other retrieval methods use graph-based knowledge along with transformer models to find multi-hop reasoning paths \cite{saxena2020improving}\cite{wang2018attention}.

\begin{figure*}[ht]
    \centering
    \includegraphics[width=\linewidth,trim=0cm 0.68cm 0cm 0.45cm,clip]{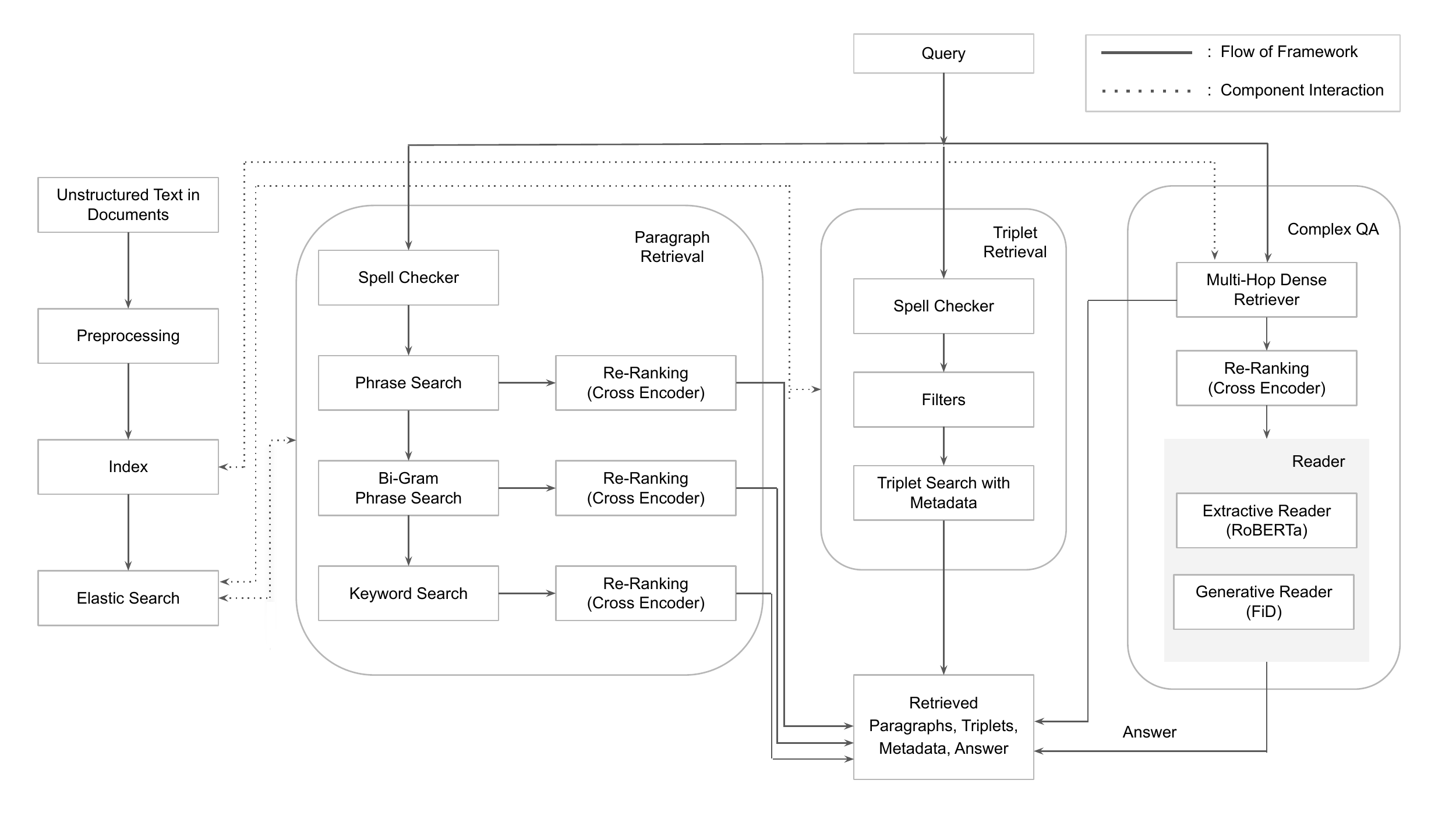}
    \caption{Architecture of Complex Information Retrieval System. The query is passed through all three components of the framework. The paragraph retrieval combines results from the phrase, bigram, and keyword searches and retrieves relevant passages from the indexed data. The triplet retrieval retrieves related subject-object-relation pairs from the constructed knowledge graph. The complex question answering system gives an answer to the query along with the semantically retrieved passages from the Multi-hop Dense Retriever (MDR). 
}
\end{figure*}

\section{Methodology}

In this section, we explain the proposed model architecture. We use CORD-19 \cite{wang2020cord} as an example dataset for explaining the pipeline and process throughout this paper, although the entire framework is flexible and should be translatable to a variety of datasets in biomedical literature.

\begin{figure*}[t]
    \centering
    \includegraphics[width=\linewidth,trim=0.2cm 14.5cm 3cm 0cm,clip]{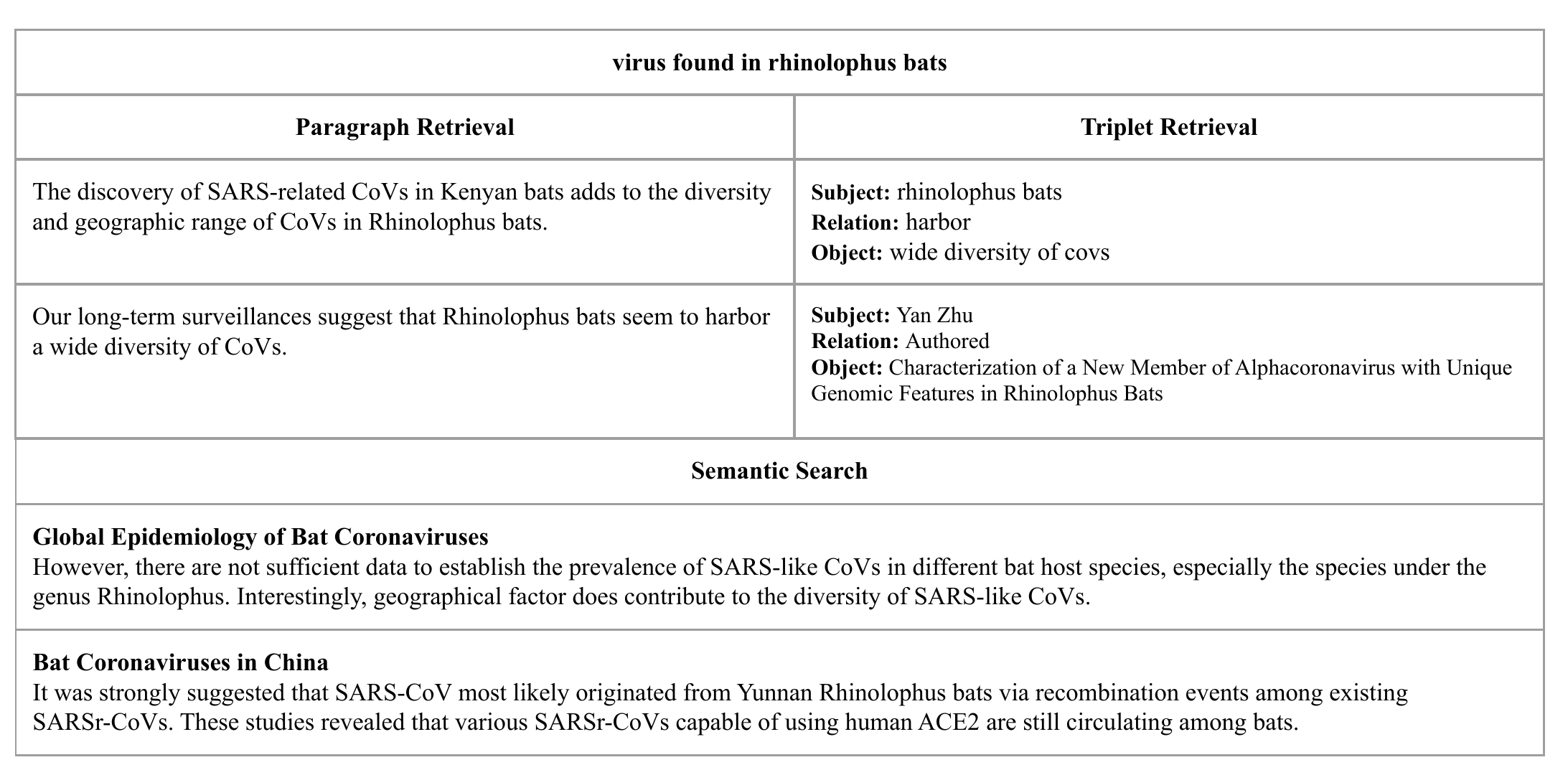}
    \caption{Results from Complex Information Retrieval Framework for phrase. The paragraph retrieval retrieves passages relevant to the detection of rhinolophus bats. The triplet retrieval results subject-relation-object pairs from the knowledge graph. They include entity-relation-entity triplets from the passages and metadata triplets like document-reference-documents. The semantic search results contain passages retrieved from Multihop Dense Retriever (MDR). As the query is a phrase, there is no response from the question-answering pipeline.}
    
\end{figure*}

\subsection{Knowledge Synthesis}
In the biomedical domain, data may be present in the form of blogs, articles, research papers, clinical documents, etc. We adopt a knowledge synthesis process to extract information in the form of subject-relation-object triplets from such unstructured text documents.

\paragraph{Knowledge graph construction}
We first clean and preprocess the text from research documents. This data is indexed for further use by the retrievers to retrieve relevant contexts. Then we pass this text through our knowledge graph construction pipeline, which is as follows:-

\begin{enumerate}
    \item Coreference Resolution on the sentences.
    \item Extracting triplets (subject, relation, object pairs) using the Open Information Extraction (OpenIE 6) System  \cite{kolluru2020openie6} from sentences.
    \item Canonicalization of the extracted relations.
    \item Linking extracted entities to appropriate ontology.
\end{enumerate}

The above pipeline results in the formation of a knowledge graph, which consists of canonicalized and linked triplets, extracted from the biomedical documents. We also include metadata such as Authors, Institutions, Publication Year, etc. along with textual phrases from the original documents.


\subsection{Complex Information Retrieval}
Our Complex Information Retrieval system consists of three main components:- \textit{Paragraph Retrieval}, \textit{Triplet Retrieval}, and \textit{Complex Question Answering}. We also have a spell checker that corrects spelling errors in the query asked by the user.

\textit{Spell correction:} Often queries include misspelled terms resulting in irrelevant results. Therefore, a spell correction module trained on biomedical text is deployed to correct the query before handing it over to the retrievers, enabling the system to handle adversarial examples of misspelled terms robustly. The module is based on \cite{norvig2007write}, which uses Levenshtein Distance (edit distance) and the probability of the word appearing in the document. 
\\

\begin{equation}
 correction(w) = argmax_{c \in candidates} P(c|w)
\end{equation}

Out of all possible candidate corrections, having an edit distance of 2 or less, the algorithm finds the correction c that maximizes the probability that c is the intended correction, given the original word w.

\subsubsection{Paragraph Retrieval}
To retrieve the most relevant piece of information from indexed documents, we introduce the paragraph retrieval functionality, where one paragraph is considered a unit of information and indexed. The retrieval combines four different search mechanisms viz, phrase search, bi-gram search, keyword search and semantic search. For all the mechanisms, we use ElasticSearch for indexing. We employ a cross-encoder to re-rank the results based on their relevance to the given query.

\paragraph{Phrase Search}
Phrase search finds an exact match for the entire query or a part of the query, which can be specified by encoding the phrase in double quotes, e.g. \textit{Human “SARS-CoV” infection} where we retrieve the relevant documents by matching the exact phrase \textit{“SARS-CoV”}. 

\paragraph{Bi-gram Search}
Bi-gram search splits the query into pairs of words, called bi-grams. These bi-grams are substrings of the query. The system searches the paragraph corpus for exact matches of these bi-grams (e.g. \textit{Real-time PCR assay} > [\textit{‘Real-time PCR’, “PCR assay’}]).

\paragraph{Keyword Search}
This method tokenizes the query and searches through the corpus for matches and retrieves them in order of the count of matches in the specific paragraph. We use an Edge n-gram tokenizer with \textit{n} being set to a minimum value of 4 and a maximum value of 30. The similarity function we use in this method is Okapi BM25 \cite{jones2000probabilistic}.

\textit{Re-ranking the results}
To re-rank the retrieved results based on relevance to the query, we use a MiniLM cross-encoder \cite{wang2020minilm} trained on MS MARCO \cite{nguyen2016ms} is used. This model outputs a relevancy score between 0 and 1 for every paragraph paired with the query. The order is decided based on this score with 1 being the highest.

\textit{Retrieved Paragraphs}
The results list consists of a predefined number of paragraphs (\textit{r}). The passages are retrieved using phrase, bi-gram, and keyword search, in the respective order. This ordering is based on descending precision for individual mechanisms. We combine these passages with the passages retrieved using semantic search, from the retriever of the complex QA system. The retrieval process continues until the length of the results list is less than \textit{r} (e.g. \textit{r}=20).

\subsubsection{Triplet Retrieval}
To retrieve the most relevant triplets for the query from our large knowledge graph we need the triplet retrieval system. This methodology retrieves triplets (subject-relation-object) which are constructed using the Knowledge Synthesis pipeline (Section III.A) for all the CORD-19 research papers. An additional feature of faceted refinement is added on top to refine the results further by specifying values for different facets. We consider the subject and object types and subtypes as facets and join multiple such facets using a boolean AND condition to filter the retrieved results.

\paragraph{ Triplet Index Construction}
While indexing the data into the ElasticServer we use the following custom settings and analyzer for preprocessing the raw JSON data:- 
\begin{enumerate}
  \item Tokenize the documents using the Edge n\_gram method.
  \item Filter the tokens to lowercase and ASCII folding.
\end{enumerate}

\paragraph{Retrieval}
The triplet retrieval component consists of the same similarity functions and search mechanisms used before viz. phrase search, bi-gram search, and keyword search. The results here consist of a list of triplets each containing subject, relation, and object. Additionally, we utilize triplet metadata like aliases, types, subtypes, descriptions, etc. Higher weightage is given to the subject, object, and relation triplet as compared to the metadata. Here, the weights can be manually tuned or trained.

\paragraph{Faceted Refinement}
Faceted refinement is employed to assist researchers to refine the information retrieved using the facet fields as shown in Fig.3. Subject and object types and subtypes are considered  facets. Multiple facets are joined together using a boolean AND condition filtering the retrieved results.

\paragraph{Knowledge Graph Querying}
We also store our knowledge graph in the Neo4j graph database AuraDB with a particular schema to run structured queries on top of it for retrieving triplets and subgraphs, using CypherQL \cite{10.1145/3183713.3190657}.


\subsubsection{Complex Question Answering}
The Complex Question Answering system can handle factoid questions, e.g. \textit{"Where was coronavirus first discovered?"}  as well as multi-hop questions which require going through multiple passages to answer the question, e.g. \textit{"What bats are the main reservoir of the virus which is transmitted to humans via ACE2 receptor?"}. We split the passages from COVID-19 related documents into a maximum length of 300 tokens. Then, we pass these fixed length passages through a transformer encoder to generate dense embeddings for each passage. We store these embeddings in a dense index for further retrieval.

\paragraph{Retriever}
The retriever searches through the dense index of CORD-19 documents and retrieves passages relevant to the query. To deal with multi-hop questions, we make use of the Multi-hop Dense Retriever (MDR) \cite{xiong2020answering}, which is an iterative retriever that uses a single RoBERTa-base model \cite{liu2019roberta} to encode queries and passages into the same vector space. It is trained to iteratively search and retrieve relevant documents from the database using Facebook AI Similarity Search (FAISS) \cite{johnson2019billion}. We have set the number of iterations to 2 in our system, but it is tunable. MDR retrieves two passages, related to each other based on reasoning paths or information about the entities in question, constituting one chain of retrieved contexts. Top-k such chains are retrieved based on their semantic similarity scores. The chains are then sorted based on the combined similarity score of the hops and further re-rank the retrieved passages using the MiniLM cross-encoder. Then we send the passages to the reader models to generate answers. We also merge these semantically retrieved results from the MDR with the results from paragraph retrieval (Section III.B.1).

\paragraph{Reader}
The reader is responsible for providing an answer given a context. We use two readers in our framework: Extractive reader and Generative reader. Extractive readers extract continuous answer spans from the retrieved passage. In contrast, generative readers are capable of generating answers even though they may not find them in the context provided. For extractive QA we use the RoBERTa model. This model
for question answering takes the question tokens and context
tokens as inputs and predicts the answer start and end tokens. For generative reader we use the Fusion-in-decoder (FiD) \cite{izacard2020leveraging}.

\begin{figure*}[ht]
    \centering
\includegraphics[width=\linewidth,trim=0.25cm 14.5cm 3cm 0cm,clip]{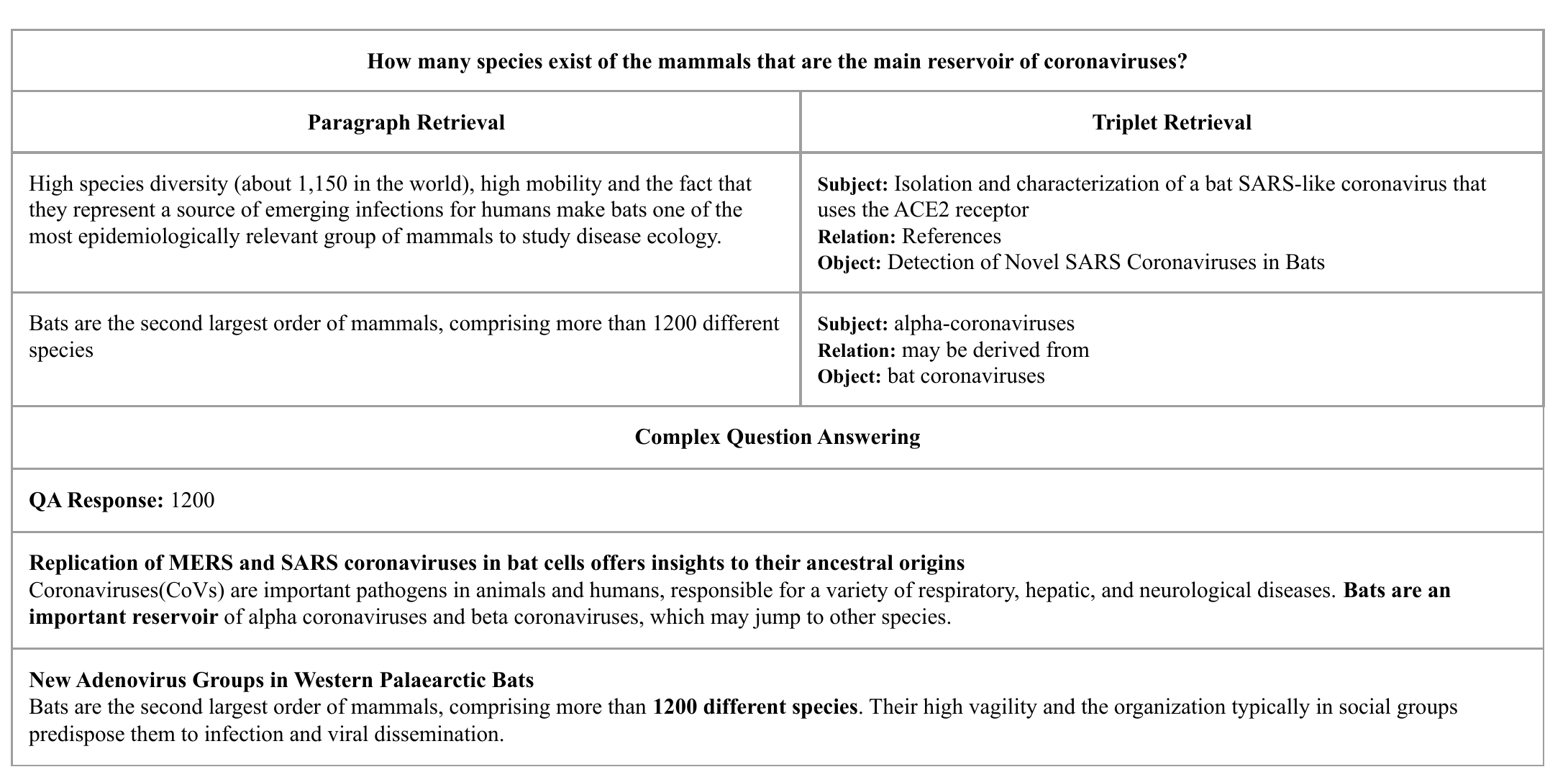}
    \caption{Results from Complex Information Retrieval Framework for a multi-hop question. The paragraph retrieval retrieves passages relevant to the question. The triplet retrieval results subject-relation-object pairs from the knowledge graph. They include entity-relation-entity triplets from the passages and metadata triplets like document-reference-documents. The complex question answering system first retrieves a passage that talks about the main reservoir of coronavirus i.e bat and then retrieves a passage that talks about the number of species of bats.}
\end{figure*}

\section{Experiments}
\subsection{Dataset}
 CORD-19 is a corpus of academic papers about COVID-19 and related coronavirus research, curated and maintained by the Allen Institute for AI. The dataset has grown to index over 1M papers and includes full-text content for nearly 370K papers. Documents from CORD-19 are indexed and information retrieval is done on top of this index. The reader models are fine-tuned on the MRQA \cite{fisch2019mrqa}  dataset that contains preprocessed subsets of other domain-related datasets, making it a more generalized and suitable benchmark. The reader is also fine-tuned on Covid-QA \cite{moller2020covid}, a medical question answering dataset around COVID-19.
 
\subsection{Training}
The extractive reader is a RoBERTa-base model, already pre-trained on WikiMultiHop. We initially fine-tune it on a generalized dataset, MRQA, and then fine-tune it further on Covid-QA for two epochs to learn the biomedical context. To avoid losing important information, we split the documents present in the CORD-19 dataset into chunks of size C, such that each chunk contains strides (overlap) of size S with the previous chunk. We make sure that C is less than 512, as most transformer models cannot process tokens more than 512 and S is set as 128 to overlap optimal information.
\\
The generative reader is the Fusion-in-Decoder model, with T5-base architecture, already pre-trained on TriviaQA \cite{joshi2017triviaqa}. We fine-tune FiD on MRQA and then on Covid-QA, for a total of 45000 steps with a batch size of 8.

\subsection{Results}
We evaluate our framework qualitatively on the CORD-19 dataset. We use two kinds of queries to test the performance of various components in our framework. First, for the phrase-\textit{“virus found in rhinolophus bats”}, we get a list of passages from paragraph retriever and multi-hop dense retriever along with multiple triplets that talk about rhinolophus bats (as shown in Fig.2). In case of a complex question like-\textit{"How many species exist of the mammals that are the main reservoir of coronaviruses?"}, the complex QA system reasons over the passages retrieved by our multi-hop retriever and the reader gives us the correct answer. It first retrieves a passage that talks about the main reservoir of coronavirus i.e, \textit{bats}, followed by a passage that talks about the number of species of bats (\textit{1200}), which can be seen in Fig.3. We also  observe that our passage retrieval mechanism retrieves highly relevant passages. They contain the keywords in the query and are contextually similar to the query asked. The triplet retrieval also retrieves the best set of triplets related to the query. Overall our system can provide the user with the most relevant information to the query asked using  lexical as well semantic retrievers unlike similar information extraction systems around COVID-19, such as \cite{su2020caire} that uses only BM25 for retrieval and not an iterative retriever like ours that also enables our question answering system to reason over more than one document and provide the answer. \cite{ambavi2020covidexplorer} supports keyword and entity search, it fails to accommodate phrase search, bi-gram search and semantic search like our search system. Both of these systems do not perform triplet retrieval on knowledge graphs.

\subsection{Evaluation}
We evaluate our framework on related open-source datasets due to the unavailability of labeled data for CORD-19. We evaluate the paragraph retrieval pipeline on another COVID-19 related dataset, TREC-COVID \cite{voorhees2021trec}. Here we use Precision and NDCG as the metric. NDCG is the ratio of the Discounted Cumulative Gain (DCG) of a recommended and ideal order. It is evident that phrase search with MiniLM-L-6-v-2 re-ranker yields better results when compared to results without re-ranking, as shown in Fig.4.
\\
We evaluate the performance of the reader models on the MRQA-dev data split by calculating the exact match and the F1 scores for all subsets of the dataset. We see that the model's performance varies massively depending on the kind of data as seen in Table.1.

\begin{figure}[H]
    \centering
    \includegraphics[width=\linewidth,trim=0cm 0.75cm 0cm 0.75cm,clip]{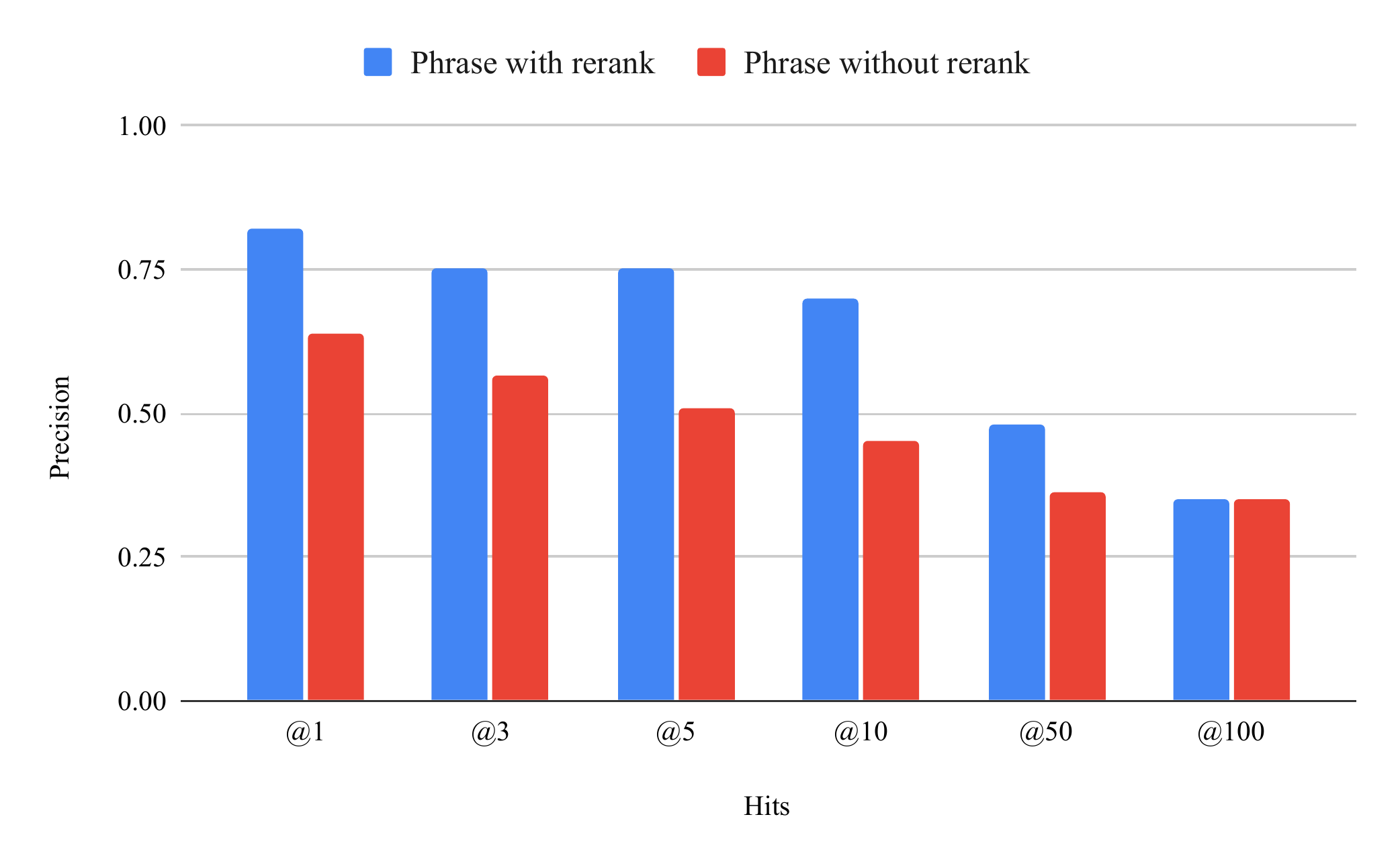}
    
\end{figure}

\begin{figure}[H]
    \centering
    \includegraphics[width=\linewidth,trim=0cm 0.75cm 1cm 0.8cm,clip]{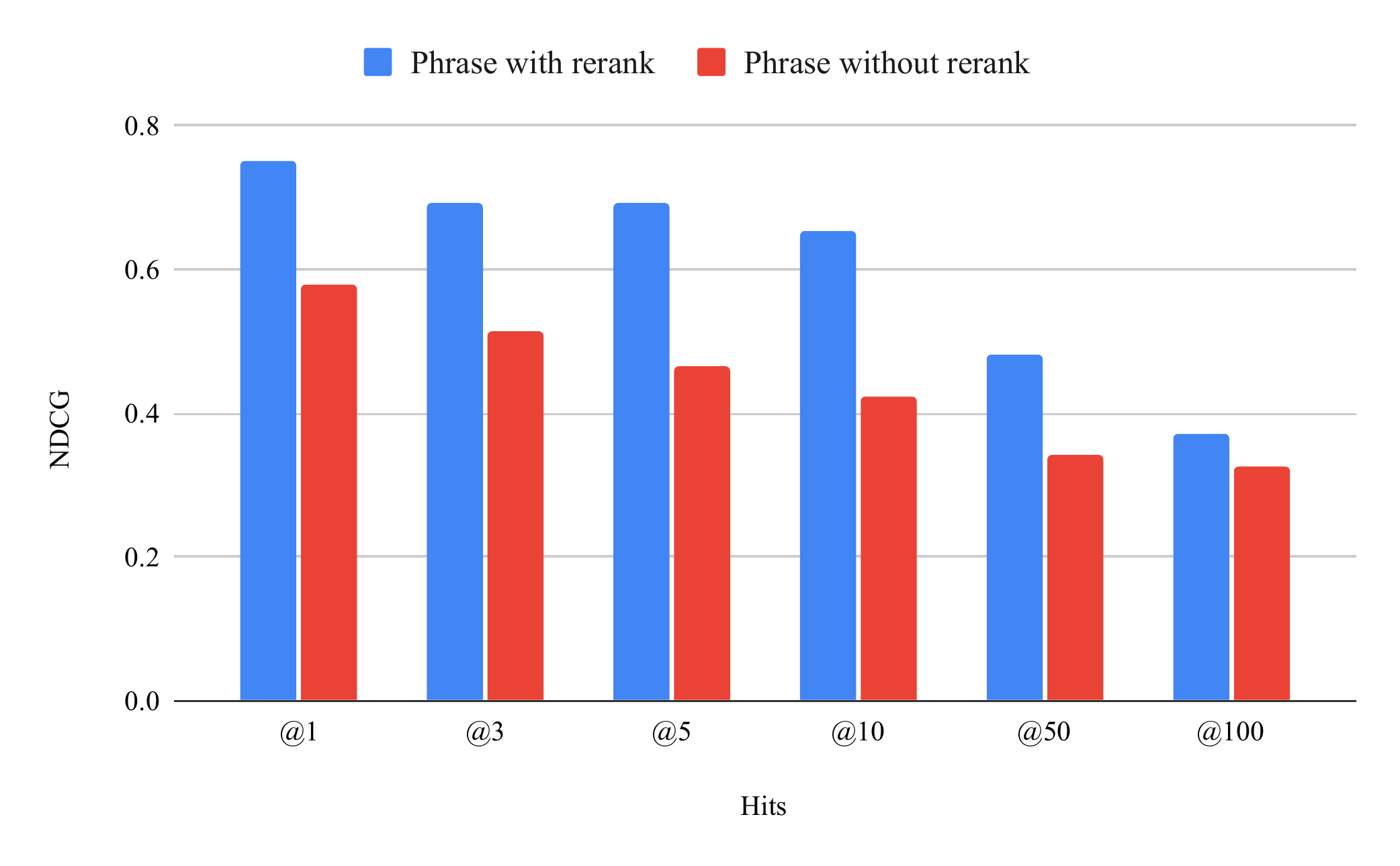}
    \caption{Evaluation of Paragraph Retrieval on TREC-COVID. Phrase with re-ranker (denoted by blue) outperforms phrase without re-ranker (denoted by red) across different top-k comparisons in both Precision and NDCG metrics.}
    
\end{figure}

\begin{table}[h]
\caption{Evaluation of Readers on MRQA-dev subsets}
\resizebox{\columnwidth}{!}{%
\begin{tabular}{lllll}
\hline
\multirow{2}{*}{Subset} &
  \multirow{2}{*}{No. of Questions} &
  \multicolumn{2}{l}{Extractive Reader} &
  Generative Reader \\ \cline{3-5} 
 &
   &
  Exact Match (\%) &
  \begin{tabular}[c]{@{}l@{}}F1\\ score (\%)\end{tabular} &
  \begin{tabular}[c]{@{}l@{}}Exact Match\\ (\%)\end{tabular} \\ \hline
SQUAD         & 10507 & 83.76 & 90.48 & 69.8 \\ 
Trivia-QA-web & 7785  & 12.76 & 14.24 & 45.7 \\ 
Search QA     & 16980 & 10.2  & 10.94 & 60.4 \\ 
Hotpot QA     & 5901  & 60.78 & 76.74 & 41.5 \\ 
NQ Short      & 12836 & 64.78 & 76.74 & 48.9 \\ 
News QA       & 4212  & 52.84 & 66.62 & 36.2 \\ \hline
\end{tabular}%
}
\end{table}

\section{Conclusion}
In this paper, we presented a complex information retrieval framework built on COVID-19 related biomedical documents that can perform both lexical and semantic search and retrieve paragraphs along with a knowledge graph consisting of triplets extracted from unstructured text. We also use faceted refinement to filter the results. We demonstrate our complex QA system, which gives the researcher a pinpoint answer to the query asked. We find that this framework makes it easier for the researcher to search for specific information from massive corpora. In our future work, we plan to add functionalities like query expansion and query intent classification along with scalable semantic retrieval on top of the knowledge graph.

\section*{Acknowledgment}
We thank Varun V, Advaith Shankar, Nim Sherpa and Saisubramaniam Gopalakrishnan for their assistance with figures, and useful suggestions that helped improve the manuscript.





\bibliographystyle{IEEEtran}
\bibliography{bib}

\end{document}